\documentstyle{article}
 
\def\beg{\begin{equation}}
\def\ene{\end{equation}}

\begin{document}
\title{\bf Modified Laplace transformation method at finite temperature: 
application to infra-red problems of\\ $N$ component $\phi^4$ theory}
\author{{\sc H. Yamada}\\e-mail: yamadah@cc.it-chiba.ac.jp\\Natural
Science, General Education\\
Chiba Institute of Technology\\Shibazono 2-1-1\\Narashino, Chiba 275\\Japan}

\maketitle
\large
\baselineskip 18pt
\vskip 3cm
\centerline{\bf ABSTRACT}

{\large Modified Laplace transformation method is applied to $N$ component
$\phi^4$ theory and the finite temperature problem in the massless limit is
re-examined in the large $N$ limit.  We perform perturbation expansion of the
dressed thermal mass in the massive case to several orders and try the massless
approximation with the help of modified Laplace transformation.  The
contribution with fractional power of the coupling constant is recovered from
the truncated massive series.  The use of
inverse Laplace transformation with respect to the mass square is crucial in
evaluating the coefficients of fractional power terms. }  

\newpage
\large
\baselineskip 24pt
\section{Introduction}
It is well known that the conventional perturbation theory for bosons breaks
down when the temperature is sufficiently high compared to the mass
scale of the theory under consideration$^{1}$.  The origin is easily seen
in the massless limit as the emergence of infra-red divergences of 
Feynman diagrams.  This is linked to the fact that the perturbative
infra-red structure
becomes worse at finite temperature.  For example, the boson propagator
shows the behavior more singular at small momentum
region than at zero
temperature, and the 
temperature correction remains finite for the self-energy while at zero
temperature it vanishes at zero mass in dimensional regularization.  The
nonvanishing of mass correction activates the hidden infra-red divergence of
sub-diagrams coupled to the tadpole, and thus the divergence
 manifests itself in Feynman amplitudes.   

The presence of the thermal mass means that there would be no infra-red problem
if one can incorporate such effect from the outset of perturbation expansion. 
It also means, as well known, that the ordinary perturbation expansion in
powers of
the coupling constant breaks down.  For example, in a single component 
$\phi^4$ theory with $\lambda\phi^4$ interaction, the pressure $P$ is given at
temperature $T\hskip 3pt(=1/\beta)$ by
\beg
P=T^4\biggl[ {\pi^2 \over 90}-{1 \over 48}\lambda+{1 \over
12\pi}\lambda^{3/2}+O(\lambda^2)\biggl].
\ene
The presence of the 
fractional power of the coupling constant $\lambda$ exhibits the break down
of perturbation expansion.  

To obtain the amplitude free from the infra-red singularity, one may sum up
a set of Feynman diagrams to all orders.  Similar but more systematic approach
 would be to shift the perturbative vacuum such that
the fields are thermally screened at the tree level$^{2,3,4,5}$.  This
program is
implemented by adding the thermal mass to be induced by the loop correction
to the
free part and subtracting from the interaction part.  Although this is
physically appealing, it is not effective in the study of magnetic part of
gluon self-energy, because the one-loop contribution is absent and higher
order diagrams are plagued with the infra-red singularity.  This is a
reason why we look for another expansion scheme.

Recently a new perturbative scheme was proposed$^{6,7}$ and it seems to be a
convenient scheme to deal with the infra-red problem at finite temperature. 
The scheme is based
on the use of the Laplace transformation which is modified as to
fit the perturbative framework.  In the case of anharmonic oscillator, the
infra-red problem at zero temperature is well controlled by the modified
Laplace transformation (MLT) method$^{7}$.  The purpose of the
present paper is to apply the method to the finite temperature perturbation
theory of
$N$ component $\lambda\phi^4$ theory at $D=4$ $^{8}$.  To check our result we
confine ourselves with the leading order of $1/N$ expansion.  We consider the
self-energy and investigate how the non-analytic structure in the
coupling constant similar to (1) can be recovered from massive perturbation
expansion which gives a simple power series in
the coupling constant.  We do not sum up Feynman diagrams to all orders,
but just deal with the truncated series.  Nevertheless we will show that
first several
coefficients of the fractional power terms can be obtained from the
truncated perturbative series.

\section{Survey of the modified Laplace transformation method}
For the self-containedness we here survey the basic points of MLT
method$^{6,7}$.  Since we confine ourselves with the approximation of a massless
theory, our presentation emphasizes the relevant respects only.

Let $f(\sigma)$ be a physical function and $\sigma$ be monotonic with
respect to the mass.  Then the massless limit is equivalent with the limit,
$\sigma\to 0$.  Suppose that $f(\sigma)$ has perturbative expansion in
$1/\sigma$ and is given at the perturbative order $k$ as
$f_{k}(\sigma)=\sigma^{p}\sum^{k}_{n=0}a_{n}/\sigma^k$ where $p$ is some
constant.  Then the "infra-red problem" occurs when we ask the
behavior of
$f(\sigma)$ for small $\sigma$.  This is the situation we encounter at
finite temperature $\phi^4$ theory.  Since we cannot take the massless
limit in $f_{k}(\sigma)$, we may fix $\sigma$ at some small value where the
lower limit of perturbation series is signaled.  In this approximation
scheme, it is convenient to consider the inverse Laplace or Heaviside
transform with respect to $\sigma$ as we can see below.
  
Heaviside
transform of $f(\sigma)$ is given by the Bromwich integral,
\beg 
\hat f(x)=\int^{s+i\infty}_{s-i\infty}{d\sigma \over 2\pi i}
\exp(\sigma x){1 \over \sigma}f(\sigma),
\ene
where the parameter $s$ represents the location of the vertical contour.  
The contour of integration should be placed on the right side to all the
possible
poles and the
cut of $f(\sigma)/\sigma$.  Then, if $x<0$, the contour may be
closed into the right half circle and $\hat f(x)$ is found to vanish.  As well
known, $\hat f(x)$ is the kernel of Laplace transformation of $f(\sigma)$.  We
have
\begin{equation}
f(\sigma)=\sigma\int^{\infty}_{-\infty}dx\exp(-\sigma x)\hat f(x).
\end{equation}
Since $\hat f(x)=0$ when $x<0$, the integration range reduces to
$[0,\infty)$. 
However $\hat f(x)$ involves the step and $\delta$ functions and it is
generally 
convenient to keep the range as $(-\infty,+\infty)$ to handle
 integration easily.

In our
massless approximation, we begin with the observation that
\beg
\lim_{\sigma\to 0}f(\sigma)=\lim_{x\to \infty}\hat f(x),
\ene
where we assumed the existence of the limits.  This ensures us that $\hat
f(x)$ can be equally used for the massless approximation.  Let us denote
the Heaviside transform of $f_{k}(\sigma)$ as $\hat f_{k}(x)$.  Although we
cannot take the massless limit for $\hat f_{k}(x)$, we may approximate
$\hat f(\infty)$ by fixing $x$ some large value, $x^{*}$, in $\hat
f_{k}(x)$.  Then we note that, in some cases including our finite
temperature problem, the transformed function has an advantage that the
convergence radius is larger than $f(\sigma)$.  For example, when $f=\sum
a_{n}/\sigma^{n}$, $\hat f=\sum
(a_{n}/n!)x^{n}$.  Therefore, if $\lim_{x\to \infty}\hat f(x)$
converges, then the transformed function gives us more chance to know the
massless 
value by increasing the order of expansion. 

To summarize, calculating $\hat
f_{k}(x)$ from (2), we approximate $f(0)$ by
$\hat f_{k}(x^*)$.  The explicit way of fixing $x^*$ will be discussed in
the next
section.  

This approach can be naturally extended to the calculation of the massive
case$^{7}$.  Let
$f_{k}(\sigma,x^*)$ be the $k$-th order approximation of $f(\sigma)$.  Then
$f_{k}(\sigma,x^*)$ is given by
\beg
f_{k}(\sigma,x^{*})=\exp(-\sigma x^{*})\hat
f_{k}(x^{*})+\sigma\int^{x^{*}}_{-\infty}dx
\exp(-\sigma x)\hat f_{k}(x).
\ene
The right-hand-side defines our modification of Laplace transformation.  In
addition to the naive cut-off term (the second term), we are led to add the
first term.  This term is critical in approximating $f(\sigma)$ for small
$\sigma$.

In performing Heaviside transformation, the following formula is useful;
\beg
\sigma^{p}\rightarrow {x^{-p} \over \Gamma(1-p)}\theta(x) \quad (p\le 1),
\ene
where $\theta(x)$ denotes the step function ($\theta(x)=0$ for $x<0$ and $1$
for $x>0$).  
For larger $p$ one can obtain the transform by using
\beg
\sigma f(\sigma)\rightarrow {\partial \hat f(x) \over \partial x}.
\ene
For example we have
\beg
\sigma^{1+p}\rightarrow 
-p{x^{-1-p} \over \Gamma(1-p)}\theta(x)+
{x^{-p} \over \Gamma(1-p)}\delta(x).
\ene
In this way we find
\beg
\sigma^{n}\rightarrow 
\left\{
\begin{array}{@{\,}ll}
1 & (n=0)\\
0 & (n=1,2,3,\cdots)
\end{array}\right. 
\ene
where we have omitted the $\theta$ and $\delta$ functions.  
We note that for fractional $p$ the transform does not vanish.

\section{Application of MLT method}
We work with the imaginary time formalism of finite temperature field theory.  
The model we consider is defined by$^{8}$:
\beg
{\cal L}_{E}={1 \over 2}(\partial \phi)^2+{1 \over 2}m^2
\phi^2+{\lambda\mu^{2\epsilon} \over 4 N}\phi^4+{\cal L}_{E,c},
\ene
where $\phi^2=\sum^{N}_{n=1}\phi_{n}^{2}$ and ${\cal L}_{E,c}$ denotes the
collection of counter terms.  
We use dimensional regularization at $D=4-2\epsilon$ with MS
prescription.  Before turning to detailed calculation, we spend some
words on our basic strategy.

We write the massive perturbation expansion of the thermal mass to order $k$
as
\beg
M^{2}=m^{2}+\sum^{k}_{n=1}c_{n}(m)g^{n}.
\ene
Since we are interested in the small mass behavior of massive perturbation
theory, we expand each loop correction in the mass.  This is equivalent to use
the high temperature expansion.  
For example, the one-loop contribution to the thermal mass gives
\begin{eqnarray}
c_{1}&=&g\biggl[{8\zeta(2) \over \beta^2}-{4\pi m \over \beta}-t m^2+{8\pi^{3/2}
\over \beta^2}\sum^{\infty}_{n=2}{(-1)^n \over n!}\Bigl({\beta^2 m^2
\over 4\pi^2}\Bigl)^n \zeta(2n-1)\Gamma(n-1/2)\biggl]\nonumber\\
&=&g\biggl[{8\zeta(2) \over \beta^2}-{4\pi m \over
\beta}+R_{even}(m^2)\biggl],
\end{eqnarray}
where
\beg
g={\lambda \over (4\pi)^2},\quad t=\gamma+\log{\beta^2 \mu^2 \over 4\pi}.
\ene
The term linear in $m$ is the crucial one in this correction.  It is in
a sense non-perturbative, since the mass enters in the Lagrangian as the
square of the mass, $m^2$.  Actually all the infra-red divergent behavior 
of sub-diagrams with many $\phi$ legs comes from that term by the
differentiation with respect to $m^2$.  
 
There is a subtlety in choosing the transformation variable $\sigma$.  For
instance we can enlarge the convergence radius of the thermal mass,
$M^2(m)$, in both cases where
$\sigma=m$ and $\sigma=m^2$.  In this paper, we choose the choice,
$\sigma=m^2$, according to the successful choice in the anharmonic
oscillator$^{7}$. 

Heaviside transform of $c_{1}(m)$, $\hat c_{1}(x)$, is then given by
\beg
\hat c_{1}(x)=g\biggl[{8\zeta(2) \over \beta^2}-{4\pi \over \beta
\sqrt{x}\Gamma(1/2)}+\hat
R_{even} (x)\biggl],
\ene
where $\hat R_{even}(x)$ denotes the Heaviside transform of $R_{even}(m^2)$. 
Since any term
in $R_{even}(m^2)$ is an even positive power of $m$, any finite sum
involved gives zero
(see (9)). 
Although this
does not necessarily mean that $\hat R_{even}(x)$ is zero, we may safely
neglect it
when we focus on the massless approximation.  On the other hand, while the
second term, $4\pi/\beta \sqrt{x}\Gamma(1/2)$, tends to zero when $x\to
\infty$, we keep it because it gives non-negligible contribution at finite
perturbative order where $x$ is fixed to some finite value in the massless
approximation.  To summarize, we keep terms as the transformation suggests:
 Terms needed are of negative, zero and odd powers of $m$ (which
corresponds to positive, zero and fractional powers of $x$
). 

Before performing our approximate calculation to the full several loops, let us
illustrate in the next subsection how the leading fractional term of order
$g^{3/2}$ comes out in our
approach.

\subsection{Calculation of $g^{3/2}$ term}
It is well known that the leading fractional term of order $g^{3/2}$ comes
from the series of Feynman diagrams shown in Fig.1, each of which gives
most dominant contribution
for small $m$ at each order in $g$.  Let us first focus on the dominant
graphs and try to
approximate the leading fractional term.

The diagrams shown in Fig.1 gives:
\begin{eqnarray}
M^2_{dominant}(m)&=&{8\zeta(2) \over \beta^2}\biggl[g\Bigl(1-{3\beta m
\over \pi}\Bigl)-{2\pi \over \beta
m}g^2+{8\zeta(2)\pi \over \beta^3 m^3}{g^3 \over 2!}-{(8\zeta(2))^2\pi \over
\beta^5 m^5}{1\cdot 3  \over 3! 2}g^4\nonumber\\
&+&{(8\zeta(2))^3\pi \over \beta^7
m^7}{1\cdot 3\cdot 5  \over 4! 2^2}g^5+\cdots\biggl]
+({\rm sub\hskip 3pt dominant\hskip 3pt 
terms}),
\end{eqnarray}
where $({\rm sub\hskip 3pt dominant\hskip 3pt 
terms})$ denotes the sub-dominant pieces in the small $m$ expansion.
Hereafter we omit those sub-dominant terms.  We have kept the next-to-the
leading term (the second one in (15)) in the one-loop contribution.  This
is because the leading contribution of each higher order
diagram comes
from this term and the power counting of higher terms with respect to $m$
suggests that it is the first term in the dominant series.  
From (6) the Heaviside transform of $M^2_{dominant}$ with respect to $m^2$ is
found to give
\begin{eqnarray}
\hat M^2_{dominant}(x)&=&{8\zeta(2) \over \beta^2}\biggl[g\Bigl(1-{3\beta
\over \pi\sqrt{x}\Gamma(1/2)}\Bigl)-{2\pi \sqrt{x} \over
\beta \Gamma(3/2)}g^2+{8\zeta(2)\pi x^{3/2} \over \Gamma(5/2)\beta^3}{g^3
\over 2!}\nonumber\\
&-&{(8\zeta(2))^2\pi
x^{5/2} \over \Gamma(7/2)\beta^5}{1\cdot 3  \over 3! 2}g^4+\cdots\biggl].
\end{eqnarray}

Nontrivial result is obtained from the 2-loop order.  At 2-loop order, we have
\beg
\hat M^2_{dominant}(x)={8\zeta(2) \over \beta^2}\biggl[g\Bigl(1-{3\beta
\over \pi\sqrt{x}\Gamma(1/2)}\Bigl)-{2\pi \sqrt{x} \over
\Gamma(3/2)\beta }g^2\biggl].
\ene
The function $\hat M^2_{dominant}(x)$ behaves as shown in Fig.2 and the
limitation of the result is seen, for example for $g=0.05$, at
$x/\beta^2\sim 1.52$.  When $x/\beta^2$
exceeds $1.52$, the highest term in (17) dominates over and
therefore the result (17) loses its validity for larger $x$.  Thus, the upper 
limit of perturbative region may be given at the stationary point of $\hat
M^2_{dominant}(x)$ and it is specified by
\beg
{\partial \hat M^2_{dominant}(x) \over \partial x}={8 \zeta(2) \over
\beta^2}\biggl[{3g\beta \over
2\pi\Gamma(1/2)x^{3/2}}-{\pi g^2 \over \Gamma(3/2)\beta x^{1/2}}\biggl]=0.
\ene
We note that the first term in (17), which is independent of $x$, does not
enter the stationarity condition (18).  
The solution of (18) is given by
\beg
x^{*}={3\beta^2 \over 4\pi^2 g}.
\ene
Note that $x^{*}$ is proportional to $g^{-1}$.  Hence the fractional power
in $x$
produces
the fractional power of $g$.  This is the mechanism to generate the
fractional power
of $g$ which generally holds in our approach.   
By substituting $x^{*}$ into $\hat M^2_{dominant}(x)$ we obtain the
approximation of $M^2_{dominant}(m=0)$.  The result reads
\beg
\hat M^2_{dominant}(x^{*})={8\zeta(2) \over \beta^2}\biggl[g-{4\sqrt{3}
\over \sqrt{\pi}}g^{3/2}\biggl].
\ene
Thus, we have recovered the $g^{3/2}$ term from the truncated power series in
$g$. Since the exact result reads $M^2(m=0)=8\zeta(2)/
\beta^2\Bigl[g-2\sqrt{3}g^{3/2}+O(g^2)\Bigl]$, the first term in (20) is
exact and the error in $O(g^{3/2})$ is
about $13$ percents.  This is satisfactory as the lowest order approximation.

Now we investigate the effect of multi-loop dominant graphs. 
Since
the stationarity condition always gives $x^{*}$ as $\beta^2/(\zeta(2)g)$
times the number, it is convenient to define
\beg
\sigma={\beta^2 m^2 \over 8\zeta(2)g }.
\ene
and consider the transformation with respect to $\sigma$.  We note that
$m^2/g$ is renormalization group (RG) invariant and therefore $\sigma$ is.  In
terms of $\sigma$, (15) is written as
\beg
M^2_{dominant}={8\zeta(2) \over
\beta^2}\biggl[g-\sqrt{3}g^{3/2}A(\sigma)\biggl],
\ene
where
\beg
A(\sigma)=\sqrt{\sigma}\biggl[2+{1 \over \sigma}-{1 \over \sigma^{2}}{1 \over
2!2}+{1 \over
\sigma^{3}}{1\cdot 3
\over 3!2^2}-{1 \over \sigma^{4}}{1\cdot 3\cdot 5 \over
4!2^3}+\cdots\biggl]=\sqrt{\sigma}\biggl[2+\sum^{\infty}_{n=0}{(-1)^n \over
\sigma^{n+1}}{(2n-1)!! \over
(n+1)!2^{n}}\biggl].
\ene
Let us truncate (23) to the $k+2$ loop level.  The truncated series is given by
\beg
A_{k}(\sigma)=2\sqrt{\sigma}+\sum^{k}_{n=0}{(-1)^n \over
\sigma^{n+1/2}}{(2n-1)!!
\over
(n+1)!2^{n}}
\ene
The corresponding Heaviside function $\hat A_{k}(x)$ is given by
\beg
\hat A_{k}(x)={2 \over \Gamma(1/2)x^{1/2}}+\sum^{k}_{n=0}{(-1)^n (2n-1)!! \over
(n+1)!2^{n}\Gamma(n+3/2)}x^{n+1/2}.
\ene
Here $x$ is the conjugate variable of $\sigma$ defined by (21) and should not
be taken as the previous $x$ appeared in (16), (17), (18) and (19).   We
emphasize
that, while the convergence radius of $A(\sigma)$ is 1, that of $\hat A(x)$ is
infinite.  As mentioned in the introduction, this is the essential
advantage in dealing with $\hat A_{k}(x)$ rather than $A_{k}(\sigma)$.  
Our task is to approximate $A(0)$, which is equal with $2$, by $\hat
A_{k}(x)$ where $x$ may be optimized by the stationarity requirement.  The
stationarity condition to fix $x$ leads to $\partial 
\hat A_{k}(x)/\partial x=0$ and the solution is found at even orders.  By
substituting the solution into $\hat A_{k}(x)$, we obtain the following
approximation for $A(0)=2$;
\begin{eqnarray}
2-loop &:& 2.25676\quad(x^{*}=1),\nonumber\\
4-loop &:& 2.06054\quad(x^{*}=1.596),\nonumber\\
6-loop &: & 2.02096\quad(x^{*}=2.181),\nonumber\\
8-loop &: & 2.00835\quad(x^{*}=2.759),\nonumber\\
10-loop &:& 2.00358\quad(x^{*}=3.334). 
\end{eqnarray}
Thus, the sequence gives the good approximation of $A(0)$ and $g^{3/2}$
term.  Finally we point out that the omitting of the first term in
$A_{k}(\sigma)$ (see (24)) gives poor result and its necessity is
confirmed.

\subsection{Approximation via full loop expansion to 10-loops}
The previous procedure can be applied to the full loop corrections of the
dressed thermal mass.  

We first sum up the logarithmic corrections
by fixing the value of $t$.  Let us fix $t$ simply by
\beg
t=\gamma+\log{\beta^2\mu^2 \over 4\pi}=0.
\ene
This gives $\mu^2=4\pi e^{-\gamma}/\beta^2$ and the coupling constant 
$g$ is given by
\beg
g=\biggl[\log{\beta^2\Lambda^2 \over 4\pi}+\gamma\biggl]^{-1}\equiv g_{\beta},
\ene
where $\Lambda$ denotes the intrinsic scale in MS scheme.  
If one wishes to recover the ordinary logarithmic terms, one can do it by 
expanding $g_{\beta}$ as $g_{\beta}=g(\mu)\Bigl(1+t g(\mu)+t^2
g^2(\mu)+\cdots\Bigl)$.  In
this case, RG invariance will be satisfied to the same level of ordinary
perturbative expansion.

From 2-loop level, the terms with odd power of $m$
appear to all orders in the high temperature expansion.  For instance
\begin{eqnarray}
{\beta^2 \over 8\zeta(2)}c_{2}&=&g_{\beta}^2\biggl(6-{2\pi \over 3\beta
m}-{15\beta^3 m^3 \over
16\pi^3}\zeta(3)+{21\beta^5 m^5 \over
128\pi^5}\zeta(5)+\cdots\biggl)\nonumber\\
&=&
6 g_{\beta}^2-\sqrt{3}g_{\beta}^{3/2}\biggl[{1 \over
\sqrt{\sigma}}+{5\over 6}g_{\beta}^2\zeta(3)\sigma^{3/2}-{7 \over
36}g_{\beta}^3\zeta(5)\sigma^{5/2}+\cdots\biggl],
\end{eqnarray}
where
$\cdots$ denotes the higher order terms in $m$ and
even powers of $m$ is fully omitted.  The terms with positive and odd
powers of $m$ survive under the Heaviside transformation with respect to
$m^2$.   Hence,  although those terms go to zero as $\sigma\to 0$, they are
needed at
finite order massless approximation.  We note that any term in the small mass
expansion of $c_{2}$ is of $g_{\beta}^2$-order, but the higher order term in
$m$ in
fact contributes to the higher order in $g_{\beta}$ of massless approximant as
suggested in (29).

In small mass or high-temperature expansion, $m$ independent pieces (the
first term in (12) and (29), for example) play a
special role under our approach because those are not affected
by our approximation scheme and go through to the final result.  Those
terms are collected as
\beg
{8\zeta(2) \over \beta^2}\biggl[g_{\beta}+6 g_{\beta}^2+{1 \over
6}\zeta(3)g_{\beta}^3+O(g_{\beta}^4)\biggl]\equiv M^2_{g}
\ene
and give the purely power like part of $M^{2}(m=0)$.  Actually the above
result agrees with the exact result.  On the other hand, the terms with
fractional 
power of $\sigma$ give rise to the fractional power (in $g$) part of the
approximant of $M^2(m=0)$.  At this stage, let us summarize the
classification of terms responsible for the approximation of $M^2(m=0)$. 
The terms zero-th order in small mass expansion were gathered into
$M^2_{g}$ and exactly gives the integer power part of $M^2(m=0)$.  The
terms of odd power of $m$ are gathered into $M^2_{\sqrt{g}}$ and contribute
to the fractional power part of $M^2(m=0)$.  Thus, neglecting terms of even
positive powers of $m$ (including the tree-level contribution), we can put 
\beg
M^2(m)=M^2_{g}+M^2_{\sqrt{g}}(m).
\ene 

Now we explicitly carry out the
approximation of the massless value of $M^2_{\sqrt{g}}$.  
We have calculated the radiative corrections $c_{n}$ up to 10-loops and
classified the result in powers of $g_{\beta}$.  Then we can write 
\begin{eqnarray}
M^2_{\sqrt{g}}&=&-\sqrt{3}g_{\beta}^{3/2}f(\sigma,g_{\beta})\nonumber\\
&=&-\sqrt{3}g_{\beta}^{3/2}\biggl[A(\sigma)+g_{\beta}B(\sigma)+
g_{\beta}^2 C(\sigma)+O(g_{\beta}^3)\biggl].
\end{eqnarray}
We further classified  
coefficient functions, $A(\sigma), B(\sigma), \cdots$, based
on $\zeta$ functions.  The result is shown in appendix.  We thus arrive at
\begin{eqnarray}
M_{\sqrt{g}}^2&=&-\sqrt{3}g_{\beta}^{3/2}\biggl[ 2a+3 b g_{\beta}+\Bigl({5
c_{\zeta(3)} \over
6}\zeta(3)-{9c \over 4}\Bigl)g_{\beta}^2
+\Bigl({35 d_{\zeta(3)} \over 4}\zeta(3)-{7
d_{\zeta(5)} \over 36}\zeta(5)+{27 \over 8} d\Bigl)g_{\beta}^3\nonumber\\
& &+\Bigl({189 e_{\zeta(3)} \over 16}\zeta(3)+{7 e_{\zeta(3)^2} \over
16}\zeta(3)^2-{35 e_{\zeta(5)} \over 8}\zeta(5)+{5 e_{\zeta(7)} \over
96}\zeta(7)-{405 e \over 64}\Bigl)g_{\beta}^4+O(g_{\beta}^5) \biggl]
\end{eqnarray}

Our massless approximation goes as in the
previous subsection:  We first perform the Heaviside transformation of
$M^2(m)$ which gives $\hat M^2(x)=M^2_{g}+\hat M^2_{\sqrt{g}}(x)$.  Here
$\hat M^2_{\sqrt{g}}(x)$ is given by the series same as (33) except that
the 
coefficient functions $a(\sigma)$, $b(\sigma)$, $\cdots$ are now replaced
by the Heaviside transforms, $\hat a(x), \hat b(x), \cdots$.  Then we look
for
the stationary points of them to evaluate
their massless values.  The result is shown in Table 1.

We find that, at the 10-loop level, the first 5 coefficient functions,
$A,B,C,D$ and $E$ are approximately evaluated.  The coefficients
$a, b, c_{\zeta}, d_{\zeta}$ and $e_{\zeta}$ already reach to the
sufficient accuracy at this level.  The accuracy of $c, d$ and $e$ is
however not
so good.  The reason is that $\hat c(x),\hat d(x)$ and $\hat e(x)$ start
with $x^{3/2},x^{5/2}$ and $x^{7/2}$ terms,
respectively, and therefore need higher order contributions to roughly
reach their values at $x=\infty$.  From Table 1, we obtain the following
approximant of the fractional power part of $M^2(m=0)$,
\beg
M_{\sqrt{g}}^2
=-\sqrt{3}g_{\beta}^{3/2}\biggl[2.0036+2.9220
g_{\beta}-0.6172 g_{\beta}^2+11.2082 g_{\beta}^3
+8.7295 g_{\beta}^4+O(g_{\beta}^5)\biggl].
\ene
The exact result reads,
\beg
M_{\sqrt{g}}^2(m=0)=-\sqrt{3}g_{\beta}^{3/2}\biggl[2+3g_{\beta}-
0.8075 g_{\beta}^2+13.6914 g_{\beta}^3 
+4.0193 g_{\beta}^4+O(g_{\beta}^{5})\biggl].
\ene
In view of (34) and (35), we can say that our
approach
can produce the fractional power part of the self-energy to the
satisfactory level.

\section{Conclusion}
In the large $N$ limit of $N$ component $\phi^4$ theory, we found the
mechanism that how the integer and the fractional power terms
in the massless theory are systematically classified and generated via massive
perturbation theory.  The fractional terms comes from the odd power terms of
$m$ in the high temperature expansion.  To 10-loop order,
the terms of order
$g_{\beta}^{3/2},g_{\beta}^{5/2},g_{\beta}^{7/2},g_{\beta}^{9/2}$ and
$g_{\beta}^{11/2}$ were approximately calculated with the help of MLT
method.
  Although only the two point function is considered in this paper,
our
approach would be directly applied to the thermodynamic functions, without
referring to the mass correction. 

In our approach, we can deal with all
the fractional power terms on equal footing.  We have not performed infinite
sum which will be difficult in general.  The characteristic features in our
approach are that the evaluation of the massless values is based on the
truncated sum
which can be obtained for general theories and, even when the one-loop
contribution vanishes at $m=0$, our evaluation procedure works.  Therefore
we expect that our approach may be effective for more complicated theories.
 Long time ago, Linde observed
that 
perturbative hot QCD breaks down at and above a certain perturbative
order$^{9}$, since the magnetic gluon mass would be of order $g^2 T$ and
then the infinite number of diagrams contribute to a single
order in $g$ (See also refs. 10,1).  Our approach may provide a new
calculational framework toward the resolution of the problem.  Although the
calculation of higher order diagrams is tedious and involved, the
fundamental problems issued by Linde become harmless. 

\vskip 20pt
\Large
\begin{center}
{\bf Appendix}
\end{center}
\large
In this appendix we summarize our perturbative calculation.  The function
$f$ is represented as 
$$
f(\sigma)=A(\sigma)+g_{\beta} B(\sigma)+g_{\beta}^2 C(\sigma)+g_{\beta}^3
D(\sigma)+g_{\beta}^4 E(\sigma)+O(g_{\beta}^4).\nonumber
$$
Each coefficient function is given to 10-loops as follows:
\begin{eqnarray}
A&=&2a(\sigma),\nonumber\\
B&=&3 b(\sigma),\nonumber\\
C&=&{5\zeta(3) \over 6}c_{\zeta(3)}(\sigma)-{9 \over 4}c(\sigma),\nonumber\\
D&=&{35 \over 4}\zeta(3)d_{\zeta(3)}(\sigma)-{7\zeta(5) \over
36}d_{\zeta(5)}(\sigma)+{27
\over 8}d(\sigma), \nonumber\\
E&=&{189\zeta(3) \over 16}e_{\zeta(3)}(\sigma)+{7\zeta(3)^2 \over
16}e_{\zeta(3)^2}(\sigma)-{35\zeta(5) \over 8}e_{\zeta(5)}(\sigma)+{5\zeta(7)
\over 96}e_{\zeta(7)}(\sigma)-{405 \over 64}e(\sigma)\nonumber,
\end{eqnarray}
where
\begin{eqnarray}
a&=&\sqrt{\sigma}\biggl(1+{1 \over 2\sigma}-{1 \over 8\sigma^2}+{1 \over
16\sigma^3}-{5 \over
128\sigma^4}+{7 \over 256\sigma^5}-{21 \over
1024\sigma^{6}}+{33 \over 2048\sigma^{7}}-{429 \over 32768\sigma^8}\nonumber\\
& & +{715 \over 65536 \sigma^9}\biggl),\nonumber\\
b&=&{1 \over \sqrt{\sigma}}\biggl(1-{1 \over 2\sigma}+{3 \over
8\sigma^2}-{5 \over
16\sigma^3}+{35 \over 128 \sigma^4}-{63 \over 256\sigma^{5}}+{231 \over
1024\sigma^6}-{429 \over 2048\sigma^7}\biggl),\nonumber\\
c_{\zeta(3)}&=&\sigma^{3/2}\biggl(1+{3 \over 2\sigma}+{3 \over
8\sigma^2}-{1 \over
16\sigma^3}+{3 \over
128\sigma^4}-{3 \over 256\sigma^5}+{7 \over 1024\sigma^6}-{9 \over
2048\sigma^7}+{99 \over 32768\sigma^8}\biggl),\nonumber\\
c&=&{1 \over \sigma^{3/2}}\biggl(1-{3 \over 2\sigma}+{15 \over 8\sigma^2}-{35
\over 16\sigma^3}+{315 \over 128\sigma^4}-{693 \over
256\sigma^5}\biggl),\nonumber\\
d_{\zeta(3)}&=&\sqrt{\sigma}\biggl(1+{1 \over 2\sigma}-{1 \over
8\sigma^2}+{1 \over 16\sigma^3}-{5 \over
128\sigma^4}+{7 \over 256\sigma^5}-{21 \over
1024\sigma^{6}}\biggl),\nonumber\\
d_{\zeta(5)}&=&\sigma^{5/2}\biggl(1+{5 \over 2\sigma}+{15 \over
8\sigma^2}+{5 \over 16\sigma^3}-{5
\over
128\sigma^4}+{3 \over 256\sigma^5}-{5 \over 1024\sigma^6}+{5 \over
2048\sigma^7}-{45 \over 32768\sigma^8}\biggl),\nonumber\\
d&=&{1 \over \sigma^{5/2}}\biggl(1-{5 \over 2\sigma}+{35 \over
8\sigma^2}-{105 \over 16\sigma^3}\biggl),\nonumber\\
e_{\zeta(3)}&=&{1 \over \sqrt{\sigma}}\biggl(1-{1 \over 2\sigma}+{3 \over
8\sigma^2}-{5 \over
16\sigma^3}+{35 \over 128 \sigma^4}\biggl),\nonumber\\
e_{\zeta(3)^2}&=&\sigma^{5/2}\biggl(1+{5 \over 2\sigma}+{15 \over
8\sigma^2}+{5 \over 16\sigma^3}-{5
\over
128\sigma^4}+{3 \over 256\sigma^5}-{5 \over 1024\sigma^6}+{5 \over
2048\sigma^7}\biggl),\nonumber\\
e_{\zeta(5)}&=&\sigma^{3/2}\biggl(1+{3 \over 2\sigma}+{3 \over
8\sigma^2}-{1 \over
16\sigma^3}+{3 \over
128\sigma^4}-{3 \over 256\sigma^5}+{7 \over 1024\sigma^6}\biggl),\nonumber\\
e_{\zeta(7)}&=&\sigma^{7/2}\biggl(1+{7 \over 2\sigma}+{35 \over
8\sigma^2}+{35 \over 16\sigma^3}+{35 \over 128\sigma^4}-{7 \over
256\sigma^5}+{7 \over 1024\sigma^6}-{5 \over 2048\sigma^7}+{35 \over
32768\sigma^8}\biggl),\nonumber\\
e&=&{1 \over \sigma^{7/2}}\biggl(1-{7 \over 2\sigma}\biggl).\nonumber
\end{eqnarray}
The coefficients, $a,b,c_{\zeta},d_{\zeta}$ and some of $e_{\zeta}$ begin
to appear at the lower loops (1,2 and 3 loops).  On the other hand,
$c,d,e_{\zeta(3)}$ and $e$ appear from $4,6$ and $8$ loops and therefore we
need higher order calculation to accumulate enough perturbative
information.  

\vskip 20pt
\begin{center}
{\Large References}
\end{center}
\begin{description}
\item [{1}] J. I. Kapusta, Finite temperature field theory, chap. 3 and 8,
Cambridge monographs on mathematical physics (Cambridge Univ. Press, 1989).

\item [{2}] R. Parwani, Phys. Rev. D45 (1992) 4695.

\item [{3}] J. Frenkel, A. Saa and J. Taylor, Phys. Rev. D46 (1992) 3670.

\item [{4}] P. Arnold and C. Zhai, Phys. Rev. D50 (1994) 7603.

\item [{5}] R. Parwani and H. Singh, Phys. Rev. D51 (1995) 4518.

\item [{6}] H. Yamada,  Mod. Phys. Lett. A11 (1996) 1001;  Mod. Phys. Lett.
A11 (1996) 2793.

\item [{7}] N. Mizutani and H. Yamada, {\it Modified Laplace transformation
method and its application to the anharmonic oscillator}, physics/9702013,
to appear in Int. J. Mod. Phys. A.

\item [{8}] L. Dolan and R. Jackiw, Phys. Rev. D9 (1974) 3320.

\item [{9}] A. Linde, Phys. Lett. B96 (1980) 289.

\item [{10}] D. J. Gross, R. D. Pisarski and L. G. Yaffe, Rev. Mod. Phys.
53 (1981) 43.
\end{description}
\newpage
\centerline{\bf Table Caption}
\begin{description}
\item [{Table 1}]
\hskip 10pt The coefficient functions are evaluated via the stationarity
condition.  The number in () represents the loop level at which the
evaluation is made.
\end{description}
\vskip 10pt
\centerline{\bf Figure Captions}
\begin{description}
\item [{Fig.1}]  Feynman diagrams dominant at each order in the small mass
limit.

\item [{Fig.2}] $\hat M_{dominant}(x)\beta^2/(8\zeta(3))$ at 2-loop level
is plotted for $\beta=1$ and $g=0.05$.
\end{description} 
\vskip 20pt
\begin{center}
\begin{tabular}{lllllll} 
{coefficient} & {}  & {} & {} & {} & {} & {exact}\\ \hline
{$a$} & {1.12838(2)} & {1.03027(4)} & {1.01048(6)} &
{1.00417(8)} & {1.00179(10)} & 1\\
$b$ & 0.75253(4) & 0.892268(6) & 0.948243(8) & 0.974006(10)  &   & 1\\
$c_{\zeta(3)}$ & 0.56419(3) & 0.96975(5) & 0.99304(7) &  0.997867(9) &  & 1
\\ 
$c$ & 0.300901(6) & 0.549637(8) & 0.718568(10) &  &  & 1\\ 
$d_{\zeta(3)}$ & 1.12838(4) & {1.03027(6)} & {1.01048(8)} & 1.00417(10) &  & 1 
\\ 
$d_{\zeta(5)}$ & 1.06455(5) & 1.00847(7)  & 1.00192(9) & &  &  1\\
$d$ & 0.08597(8) & 0.25135(10) &  &   &    &  1\\
$e_{\zeta(3)}$ & 0.75253(7) & 0.892268(9) &  &  &  &  1\\
$e_{\zeta(3)^2}$ & 1.06455(6) & 1.00847(8) & 1.00192(10) &  &  &  1\\
$e_{\zeta(5)}$ & 0.56419(5) & 0.96975(7) & 0.99304(9) &  &  &  1\\
$e_{\zeta(7)}$ & 0.423142(3) & 0.619066(5) & 0.983706(7) & 0.997431(9) &  &  1\\
$e$ & 0.0191(10) &  &  &  &  &  1\\
\end{tabular}
\end{center}
\centerline{\Large Table 1}

\end{document}